\documentclass[acmsmall, screen]{acmart}

\title{Predicting Tail-Risk Escalation in IDS Alert Time Series}

\author{Ambarish Gurjar, L Jean Camp}
\affiliation{
  \institution{University of North Carolina at Charlotte}
  \city{North Carolina}
  \country{USA}
}
\email{your.email@example.com}

\begin{document}

\begin{abstract}
Network defenders face a steady stream of attacks, observed as raw Intrusion Detection System (IDS) alerts. The sheer volume of alerts demands prioritization, typically based on high-level risk classifications. This work expands the scope of risk measurement by examining alerts not only through their technical characteristics but also by examining and classifying their temporal patterns. One critical issue in responding to intrusion alerts is determining whether an alert is part of an escalating attack pattern or an opportunistic scan. To identify the former, we apply extreme-regime forecasting methods from financial modeling to IDS data. Extreme-regime forecasting is designed to identify likely future high-impact events or significant shifts in system behavior. Using these methods, we examine attack patterns by computing per-minute alert intensity, volatility, and a short-term momentum measure derived from weighted moving averages.

We evaluate the efficacy of a supervised learning model for forecasting future escalation patterns using these derived features. The trained model identifies future high-intensity attacks and demonstrates strong predictive performance, achieving approximately 91\% accuracy, 89\% recall, and 98\% precision. Our contributions provide a temporal measurement framework for identifying future high-intensity attacks and demonstrate the presence of predictive early-warning signals within the temporal structure of IDS alert streams. We describe our methods in sufficient detail to enable reproduction using other IDS datasets. In addition, we make the trained models openly available to support further research. Finally, we introduce an interpretable visualization that enables defenders to generate early predictive warnings of elevated volumetric arrival risk.

\end{abstract}

\maketitle

\section{Introduction}

Organizations connected to the Internet ecosystem face a continuous stream of opportunistic intrusions, ranging from automated scanning and probing to coordinated campaigns by sophisticated actors. From the defender’s perspective, such activities are observable only through local telemetry such as intrusion detection system (IDS) alerts.  IDS systems classify individual connections, attempted connections,  or traffic flows as benign or malicious \cite{denning,snort}. %cite some famous ids paper.
 
 This study is guided by the following research question: \emph{"Does the short-term temporal microstructure of IDS alert arrivals contain predictive information about imminent transitions into extreme high-intensity regimes that drive operational risk?"}. Answering this question is critical because operational failures in security monitoring are driven not by typical alert volumes, but by episodic surges that overwhelm analysts and automated pipelines \cite{SANSSOC2019}.

Identifying the risk of any individual attack or connection is done by a combination of the data and control plane information.  In this work, we expand this to examining the temporal structure of IDS alert streams. We quantify threat arrival risk, which is any expected increase in the rate of the type of attack. This is a complement to current ratings, and we report efficacy for the entire data set as well as for the alerts defined as low, medium, or high. These ratings are taken from Suricata logs collected over three months across a large public US university enterprise network [anonymized] (~251 million alerts). Suricata is an open-source network intrusion detection and prevention system that inspects live traffic and generates structured alerts based on signature and protocol analysis \cite{Suricata2010}.

In this study, we examine large-scale IDS telemetry to test whether the temporal microstructure of alert arrivals contains predictive information about future load.
Using three months of Suricata logs collected across a large network of a public university [anonymized], we transform raw alerts into per-minute count series and derive temporal features, including intensity, volatility, and momentum that capture short-term fluctuations and directional shifts in the rate of attack activity. These temporal indicators mirror feature sets commonly used in quantitative finance to anticipate rare, high-impact events or significant shifts in system behavior, such as periods of elevated volatility. 

Essentially, we frame the problem as extreme-regime forecasting, i.e., predicting whether future alert intensity will exceed the empirical 95th percentile within a short prediction horizon. To our knowledge, this is the first application of predictive tail conditional expectation to intrusion detection data \cite{Artzner}.  Using a supervised model based on gradient-boosted decision trees (XGBoost)\cite{chen2016xgboost}, we demonstrate that these temporal features yield highly accurate predictions of future high-intensity periods, revealing measurable buildup signals preceding alert surges.

The following are our contributions:
\begin{enumerate}
    \item \textbf{Measurement of temporal risk signals in IDS alert streams}: We provide a large-scale analysis of IDS alert arrival patterns and derive temporal features that capture the dynamics of threat activity.
    \item \textbf{Extreme-regime forecasting for operational risk}: We formulate future high-intensity episodes as a 95th-percentile extreme-event prediction problem and show that these temporal features produce strong predictive performance.
    \item \textbf{Interpretable early-warning machine-learning based visualization}: We develop a lightweight visual tool that overlays predicted tail risk on live alert intensity, as a proof of concept to illustrate that operators could recognize and act on impending volumetric alert escalation.
\end{enumerate}
These contributions offer a measurement-driven approach to quantifying and forecasting tail-risk behavior in IDS alert time series and highlight the operational value of temporal microstructure in cyber-threat telemetry.
 In addition to these contributions, we will make our model available for use under an open source license. 

\section{Motivation and Related Work}
\label{sec:related}

Modern enterprise networks deploy a growing number of security sensors and detection mechanisms to monitor increasingly complex threat surfaces. As a consequence, security operations centers (SOCs) observe continuous, high-volume streams of intrusion detection system (IDS) alerts generated by heterogeneous rules, signatures, and anomaly detectors. Prior empirical studies and practitioner surveys consistently report that the volume and variability of these alerts often exceed human triage capacity, leading to alert backlogs, delayed investigations, and missed incidents \cite{julisch2003clustering, SANSSOC2019, NIST80061}. This phenomenon (commonly referred to as alert fatigue ) has been identified as a primary operational bottleneck in contemporary security monitoring workflows.

From an operational perspective, the most consequential failures are not driven by typical background alert rates, but by episodic surges in alert volume that overwhelm analysts and automated pipelines. During such high-intensity epochs, queue buildup, context switching, and prioritization errors degrade both detection quality and response timeliness. Importantly, these overload conditions arise from the temporal dynamics of alert arrivals rather than from the semantics of individual alerts, suggesting that operational risk is fundamentally a property of the alert stream as a time-dependent process.

Across the surveyed literature, we see several gaps that motivate our current work. Firstly, while IDS research has made substantial progress in alert classification and correlation using machine learning techniques. These are designed for real-time analysis. In contrast, we focus on forecasting high volumes of intrusion attempts. The work done in modelling these alerts as a timeseries stops short of predicting transitions into extreme high-intensity regimes that are most critical for operational planning. Second, extreme-event and tail-risk forecasting methods are well-developed in fields such as finance, physics, and actuarial science, but these frameworks have not been adapted to the real-time forecasting of IDS alert volumes, despite strong conceptual parallels. Finally, usability and human-centered research have improved how analysts visualize and interpret alerts, but current tools lack continuous, quantitative early-warning signals that convey near-future threat pressure. Together, these gaps highlight the absence of an integrated framework that combines temporal modeling, tail-risk prediction, and operational usability to proactively inform SOC decision-making.

\textbf{Intrusion Detection Alert Management.}
Traditional IDS research has focused primarily on interpreting individual alerts. The essential function of an IDS is to distinguish benign from malicious connections ~\cite{khraisat2019survey, Diana2025, Gupta2023, Pinto2023}. 

One challenge of IDS detection approaches is the high volumes of alerts that burden security operations centers (SOCs). Significant research has focused on easing the analyst workload through improved alerts and interaction design \cite{SANSSOC2019,lee1999framework,lee2000adaptive,lee2001information,tavallaee2009detailed,yin2017deepids}.  Visualizations can be classified as graph-based, notation-based, matrix-based, and metaphor-based visualizations. In contrast, we provide a simple threshold that could be adjusted based on the risk tolerance of the organization. As such, it could be added to any of the current visualizations as a variable or presented as is~\cite{devendra2025state}.

The closest work to ours was a risk-based approach, which allowed the defenders to choose the most salient features, provided a hierarchical view.~\cite{itoh2006hierarchical} Rate of arrival was not one of the features considered.  Devalk et al brought forward the importance of timeliness in security. That work, like many approaches, used graph theoretical approaches ~\cite{devalk2022riverside} which can fail to scale~\cite{arendt2015ocelot,yin2005design,sharafaldin2019evaluation,ma2025examining,best20147}.
A framework for visualization of IDS did not provide a high-level mapping of the tasks of network defenders, like most works, assuming that visualization can be addressed separately from alert filtering~\cite{komlodi2004information}. A survey in 2019 of the visualization of security literature illustrated that these are not integrated with reduced alerts \cite{khraisat2019survey}.  Formal user tests are rarely implemented. Instead,  integrated usage cases and case studies of tool use informed design~\cite {sharafaldin2019evaluation}.

The core of our work is to provide a data-driven threshold. We provide a simple proof-of-concept visualization that can scale and is grounded in an innovative temporal clustering. However, the primary contribution here in terms of usability is identifying alerts indicative of attacks that are likely to increase in intensity.

Other researchers have sought to improve the classification of alerts to reduce workload. Julisch~\cite{julisch2003clustering} proposed root-cause clustering, showing that alerts sharing attributes such as destination IPs, ports, and signatures can be grouped into a smaller set of generalized causes.

Valeur et al.~\cite{valeur2004correlation} correlated low-level alerts using similarity features such as signature type, temporal proximity, host context (source and destination pairs), and prerequisite–consequence relationships derived from attack models to construct higher-level attack scenarios. Both studies show that IDS alert metadata contains meaningful structure and that projecting alerts onto well-chosen categorical dimensions can substantially reduce SOC complexity. Our work shares the goal of supporting SOC operations but addresses a different problem. Instead of reducing or correlating existing alerts, we aim to forecast tail risk in alert volume. By modeling the temporal dynamics of IDS alert arrivals, we identify forthcoming high-load periods when additional analyst capacity may be needed. The demonstrated value of using alert attributes in prior work motivates our stratification approach: we group alerts by severity, the most complete and least imbalanced categorical dimension in our dataset, to obtain coherent strata whose temporal behavior can be modeled reliably.

A broader survey of intrusion detection systems by Khraisat et al.~\cite{khraisat2019survey} highlights a persistent challenge: IDS deployments generate extremely large alert volumes with false-positive rates often exceeding 90\%. The survey reviews numerous alert-correlation, reduction, and compares machine-learning techniques~\cite{lee1999framework,lee2000adaptive,lee2001information,tavallaee2009detailed,yin2017deepids}.   Reduction and improved correlation are important both for improved accuracy and for reducing analyst burden. Our research , our classifications are based on likelihood of transition to high-frequency alerts. This motivates the methodological gap we target: the need for forecasting frameworks that summarize and anticipate alert surges by predicting whether the upcoming intensity $\lambda_{t+1}$ will exceed a high-percentile threshold.

\textbf{IDS Alert Streams as Time-Series Processes.}
Work more closely aligned with our setting treats IDS alert streams as time-dependent signals rather than static observations. Viinikka et al.~\cite{viinikka2009time} model aggregated IDS flows using adaptive filters, moving averages, and Kalman smoothing to suppress background noise. Their focus is anomaly detection and alert reduction, with time-series techniques serving as the primary analytical tool. In contrast, in our work alert aggregation is used only as a preprocessing mechanism to compute structured temporal features (intensity, volatility, and momentum), rather than as the primary modeling objective.

Harang et al.~\cite{harang2014burstiness} show that operational IDS event streams exhibit burstiness and significant temporal memory, underscoring the importance of modeling arrival processes in addition to using static feature representations. However, these works do not address the challenge of forecasting transitions into extreme high-intensity regimes, focusing instead on retrospective characterization and detection.

A substantial body of prior work models IDS alert arrivals using Poisson assumptions~\cite{ihler2006adaptive, ihler2007learning, scott2003markov, han2016flame, rauta2015probabilistic}. Such models assume memoryless arrivals with stable rates (an assumption that is frequently violated in practice). Coordinated campaigns, vulnerability disclosures, and patch announcements often induce sustained bursts and rate shifts that reflect temporal dependence rather than independent arrivals. To capture such dependence, self-exciting Hawkes processes have been proposed~\cite{hawkesog, Bessy-Roland_Boumezoued_Hillairet_2021}. These models are widely used to represent feedback-driven clustering in cyber events~\cite{Bessy-Roland_Boumezoued_Hillairet_2021}, mirroring their application in seismology~\cite{hawkesseismo}, insurance~\cite{SWISHCHUK2021107}, and financial markets~\cite{hawkesfinanc}.

Our initial effort similarly examined IDS alerts through the lens of Hawkes processes. While self-exciting dynamics were observable in limited subsets of the data, the resulting models did not generalize across severity strata nor temporal scales and proved insufficient for predictive forecasting of future extreme arrivals. This limitation motivated a shift away from fully parametric point-process models toward representations that emphasize predictive structure over generative completeness.

Recent work in time-series analysis supports this transition. Agarwal et al.~\cite{AgarwalMatrix2019} propose model-agnostic forecasting methods based on matrix estimation that avoid strong parametric assumptions while remaining robust to noise and missing data. Related work on multivariate singular spectrum analysis further demonstrates that low-rank temporal structure can be extracted from complex, nonstationary signals without explicit generative modeling~\cite{AgarwalmSSA2022}. These approaches reinforce the idea that, for large-scale operational telemetry, capturing short-horizon predictive structure may be more effective than fitting a single global stochastic process.

Finally, in terms of organizing and interpreting alert behavior by risk, machine-learning approaches have been widely used to cluster or classify attacks based on temporal and contextual features~\cite{guo2023review, Pinto2023, Gupta2023}. Our work builds on this line by reframing IDS alert streams as a forecasting problem focused explicitly on predicting transitions into extreme-intensity regimes, rather than solely detecting anomalies or modeling arrival mechanisms.

\textbf{Extreme-Event and High-Quantile Forecasting.}
Forecasting rare, high-impact events has been extensively studied across hydrology, physics, climate science, and nonlinear dynamical systems, where extreme outcomes often dominate system-level risk. Our work directly builds on these mathematical and conceptual models by treating IDS alert escalation as a tail event of an underlying temporal process. Pasche and Engelke~\cite{pasche2021eqrn} developed the Extreme Quantile Regression Network (EQRN), combining extreme-value theory with neural networks to estimate high quantiles in hydrological systems. In nonlinear physics, foundational studies of rogue waves, optical spikes, and atmospheric extremes demonstrate that rare events often arise from latent instability and nonlinear amplification~\cite{akhmediev2016opticalrogue,ansmann2013extreme,kantz2006extreme}. Recent advances further show that machine learning and reservoir computing can detect precursors to extreme transitions in chaotic systems by exploiting short-term temporal structure~\cite{bhat2022mlchaos,sapsis2018extremeprediction}. Across these domains, a recurring theme is that extreme events are not purely random outliers, but are preceded by nonlinear and transient signatures in temporal features—an observation that closely parallels alert-flood phenomena in IDS telemetry.

Finance offers a mature framework for operationalizing extreme-event prediction through Value-at-Risk (VaR) and high-quantile forecasting~\cite{duffie1997var}. Unlike hydrology, physics, and climate science, financial markets explicitly incorporate strategic human behavior. Our initial hypothesis was that the presence of intelligent adversaries (e.g., the composite operator described by Wyckoff~\cite{wyckoff1931method}) would make financial risk models particularly applicable to cybersecurity settings, where attackers may deliberately induce bursts and stress system capacity. Taleb’s Black Swan theory~\cite{taleb2007blackswan} similarly emphasizes that complex systems can exhibit rare, high-impact spikes arising from hidden fragility and nonlinear interactions, a perspective that closely mirrors adversarial exploitation.

Seminal studies by Jegadeesh and Titman~\cite{jegadeesh1993returns,jegadeesh2001profitability} show that momentum captures persistent directional pressure and short-horizon predictability in financial time series. Baltas and Kosowski~\cite{baltas2013timeseries} demonstrate that volatility estimators paired with momentum signals improve regime identification, while Cao and Copeland~\cite{cao2023regimeswitching} show that volatility–momentum interactions distinguish turbulent from calm regimes in Bayesian switching models. These insights motivate our use of analogous temporal features—alert intensity, volatility, and momentum to forecast transitions into extreme-intensity regimes in IDS alert streams.

From a systems perspective, closely related questions arise in the study of workload dynamics and response-time tails, where rare surges rather than average load determine system stability and service degradation. Prior work has developed rigorous frameworks for characterizing when heavy-tailed or bursty arrival processes induce extreme response-time behavior, and for identifying policies that are optimal with respect to tail probabilities rather than mean performance~\cite{scully2020characterizing,YuScully2024}. Although this literature focuses on scheduling and queueing systems, the underlying insight—that extreme outcomes are driven by structured tail dynamics rather than noise provides theoretical grounding for treating IDS alert escalation as a tail-risk forecasting problem rather than a point prediction task.

Taken together, while prior research has examined IDS alert correlation, stochastic arrival modeling, and extreme-event prediction in isolation, our work integrates these strands into a predictive framework that targets tail-risk escalation in IDS alert streams through a systems-oriented lens.

\section{Temporal Feature Engineering and Exploratory Data Analysis}

\subsection{Dataset Overview}

%data stratification
We analyze a 2-month (December 6th 2024 to Feb 2nd 2025) collection of 251 million Suricata IDS alerts captured from 120 distributed sensors deployed across an enterprise network of a public university in the USA. Our research was reviewed by the university Institutional Review Board (IRB) before any analysis to review for potential harm. Each alert includes a high-precision timestamp, signature metadata (severity, confidence, attack target), application-protocol identifiers (suricata.app\_proto), and MITRE ATT\&CK mappings (tactics and techniques).  Each alert row contains only the fields emitted by Suricata for that specific event; as a result, not all records populate all 39 fields, and the presence or absence of a field reflects what the IDS was able to extract from the underlying packet or flow. All identifiable information was anonymized to remove identifying information while preserving temporal and categorical structure. %Additional details on the dataset, including per-field metadata and completeness metrics, are included in Appendix~\ref{appendix:data}.

\subsection{Stratification Procedure}

To evaluate data quality and guide stratification, we computed for each
column: (1) the non-null completion rate, (2) the cardinality of distinct
values (capped for high-cardinality fields), and (3) up to five
representative sample values to aid semantic interpretation. Columns with
greater than 50\% completeness and between 2--20 unique categories were
considered suitable for forming statistically meaningful strata without
introducing sparsity.

This screening process yielded three potential stratification axes:
\begin{itemize}
    \item \textbf{Severity}: \emph{Critical}, \emph{High}, \emph{Medium}, \emph{Low}; %(\texttt{signature\_severity}): \emph{High}, \emph{Medium}, \emph{Low};
    \item \textbf{Application protocol}: HTTP, DNS, TLS, etc.; %(\texttt{suricata.app\_proto}): HTTP, DNS, TLS, etc.;
    \item \textbf{MITRE ATT\&CK tactic}: Exploit Public-Facing Application,
System Information Discover, Phishing, etc. %(\texttt{mitre\_tactic\_name}): \emph{Initial Access},
    %\emph{Command \& Control}, \emph{Exfiltration}, etc.
\end{itemize}

Although all three dimensions were initially viable, we selected
\textbf{severity} as the primary stratification axis for the remainder of the
analysis. Severity exhibited (i) the highest field completeness, (ii) the
lowest category imbalance, and (iii) the most stable temporal behavior across
sensors. In contrast, MITRE and protocol labels were substantially sparser and
highly skewed. Using these dimensions would have
resulted in unreliable or excessively sparse point-process estimates.
Severity, therefore provided the only axis that enabled consistent comparison
across sensors and time while preserving statistical power.

Each severity group is modeled as an independent univariate point process
under the assumption that alerts within a severity class share similar
stochastic properties. Descriptive per-minute statistics for the resulting
strata are reported in Table~\ref{tab:desc_stats}.

\begin{table*}[h]
\centering
\caption{Descriptive statistics of IDS alert data (3-month collection period). 
Values represent per-minute aggregated counts across alert strata.}
\label{tab:desc_stats}
\begin{tabular}{lrrrr}
\toprule
\textbf{Stratum} & \textbf{Total Alerts} & \textbf{Mean Rate} & 
\textbf{Std. Dev.} & \textbf{Max} \\
\midrule
Informational & 109{,}592{,}585 & 1306.79 & 1122.75 & 79{,}876 \\
Minor         & 48{,}522{,}386  & 578.58  & 951.44  & 8{,}119  \\
Major         & 41{,}015{,}488  & 489.07  & 1036.92 & 77{,}687 \\
Critical      & 473{,}118       & 5.64    & 14.26   & 1{,}225  \\
Unknown       & 168{,}562{,}614 & 2081.62 & 1133.09 & 42{,}787 \\
\midrule
\textbf{All Strata} & 250{,}933{,}988 & 4308.84 & 2765.57 & 104{,}720 \\
\bottomrule
\end{tabular}
\vspace{-0.3em}
\end{table*}

\subsection{Data Preparation for temporal analysis}

First, timestamps were parsed and normalized with error coercion to handle malformed entries. Records with invalid or missing timestamps were excluded (\textless 0.1\% of total). Events were then aggregated into fixed 1-minute time bins using floor-based discretization, creating a discrete-time count series suitable for point process analysis. 

To manage computational constraints during exploratory analysis, a stratified 10\% random sample was extracted while maintaining temporal ordering within each stratum. For computational efficiency in model fitting, we aggregated events into 1-minute bins, yielding a time series 
\[
N(t) : t = 1, 2, \dots, T
\]
where \( N(t) \) represents the count of alerts in the \( t^{\text{th}} \) minute. The resulting time series contained \( T = x \) bins spanning a 90-day observation period.

Given the per-minute count series \( N(t) \), we estimate the instantaneous event intensity 
as a smoothed rate function using an exponential moving average:
\[
\lambda(t) = \alpha N(t) + (1 - \alpha)\lambda(t-1)
\]
where \( \alpha \in (0,1) \) controls the memory of past observations (we use \( \alpha = 0.3 \)). 
This smoothing suppresses high-frequency noise while preserving burst structures, allowing 
the resulting intensity signal \(\lambda(t)\) to better approximate the underlying arrival rate of attacks 
in continuous time.

\begin{figure}[b]
    \centering
    \includegraphics[width=0.86\linewidth]{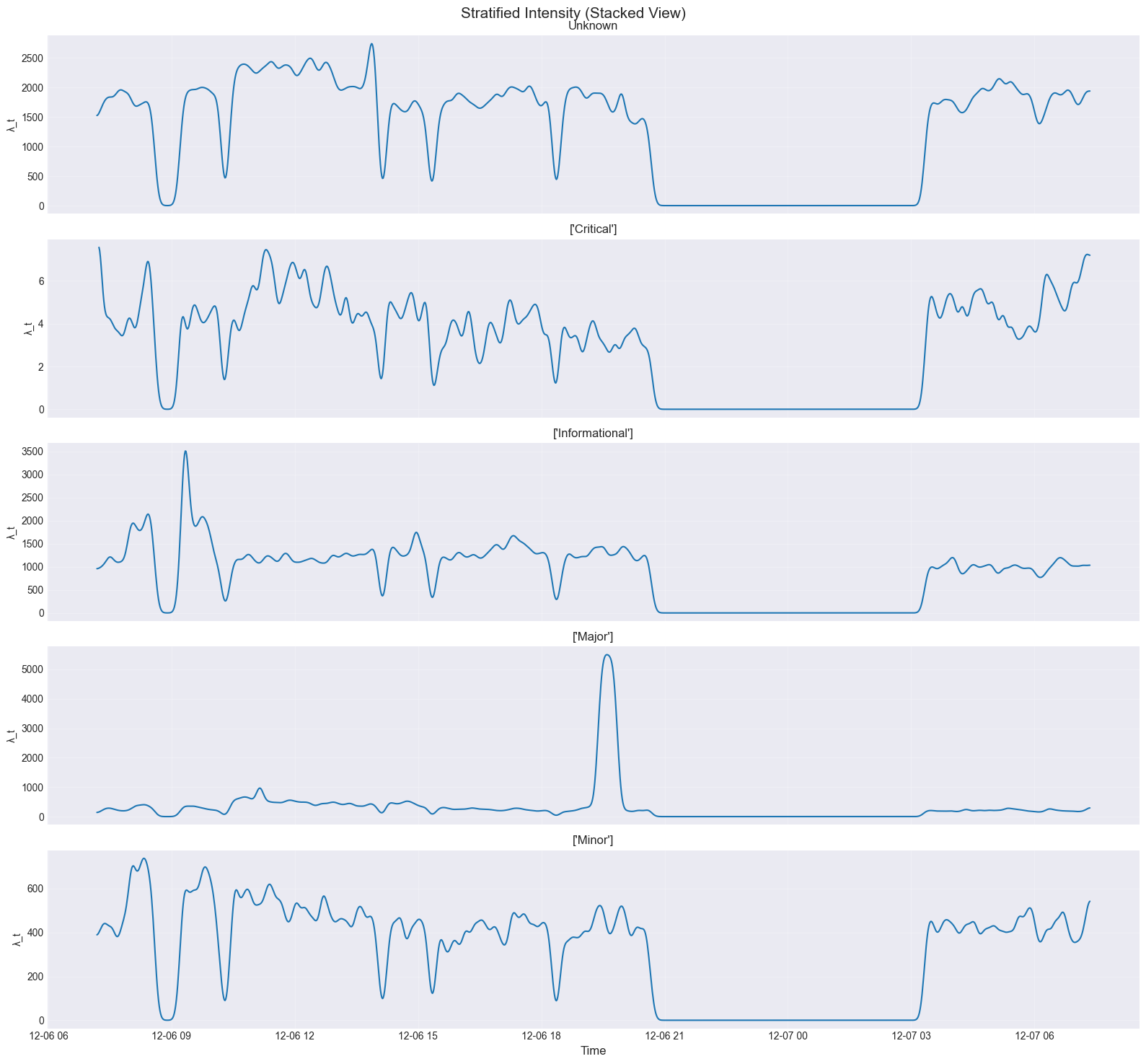}
    \caption{Temporal evolution of IDS alert intensity across severity strata. Each subplot shows the smoothed arrival intensity $\lambda_t$ for one severity class, with all subplots aligned on a common time axis for a 24-hour time slice. The night period shows a period of low activity. The plots also show heterogeneity in the temporal behavior of different severities.}
    \label{fig:stacked-intensity}

\end{figure}

\subsection{Stochastic point process Analysis}

We first modeled the IDS alert arrival process as a stochastic point process to test for the presence of self-exciting behavior in network attack activity. Using a randomly sampled 1\% subset of the data, we evaluated three canonical models: the \emph{Homogeneous Poisson process}, the \emph{Non-Homogeneous Poisson process} (NHPP), and the \emph{Hawkes process}. Goodness-of-fit was assessed via likelihood scores, time-rescaling, and Kolmogorov-Smirnov diagnostics. On the small subset, the Hawkes model exhibited the highest log-likelihood, indicating that short-term excitation is detectable at fine time scales.

However, when scaling the analysis to the full dataset, the estimated Hawkes excitation parameter diminished substantially. The global behavior more closely resembled a smoothly varying NHPP: the rate $\lambda(t)$ changed over time, but without the strong triggering structure characteristic of self-exciting processes. This discrepancy suggests that while localized burstiness exists at very short time scales, large-scale IDS dynamics are dominated by gradual fluctuations in alert volume rather than persistent cascades of excitations.

To examine how excitation varies with temporal resolution, we conducted a multiscale binning analysis using aggregation windows from 1 to 15 minutes. For each window, we re-fit the Hawkes and NHPP models and compared log-likelihoods and AIC values. The results showed that Hawkes excitation is strongest at \emph{low} aggregation scales (1--2 minutes), and rapidly weakens as the time window increases. At coarser granularities, both Hawkes and NHPP models converge in behavior, indicating that excitation signatures are largely a high-frequency phenomenon.

This multiscale behavior is operationally meaningful: defenders require early indicators of escalation, and the fine-grained scales where excitation is detectable are precisely those that matter for short-term forecasting. At the same time, these results reveal the limitations of fully parametric stochastic modeling. Hawkes models provide explanatory insight at small scales but fail to capture the broader dynamics at operational (minute-level) resolutions. This motivated our shift toward engineered temporal features that preserve short-term dynamics while remaining tractable for supervised forecasting models.

\begin{figure*}[t]
    \centering
    \includegraphics[width=0.8\textwidth]{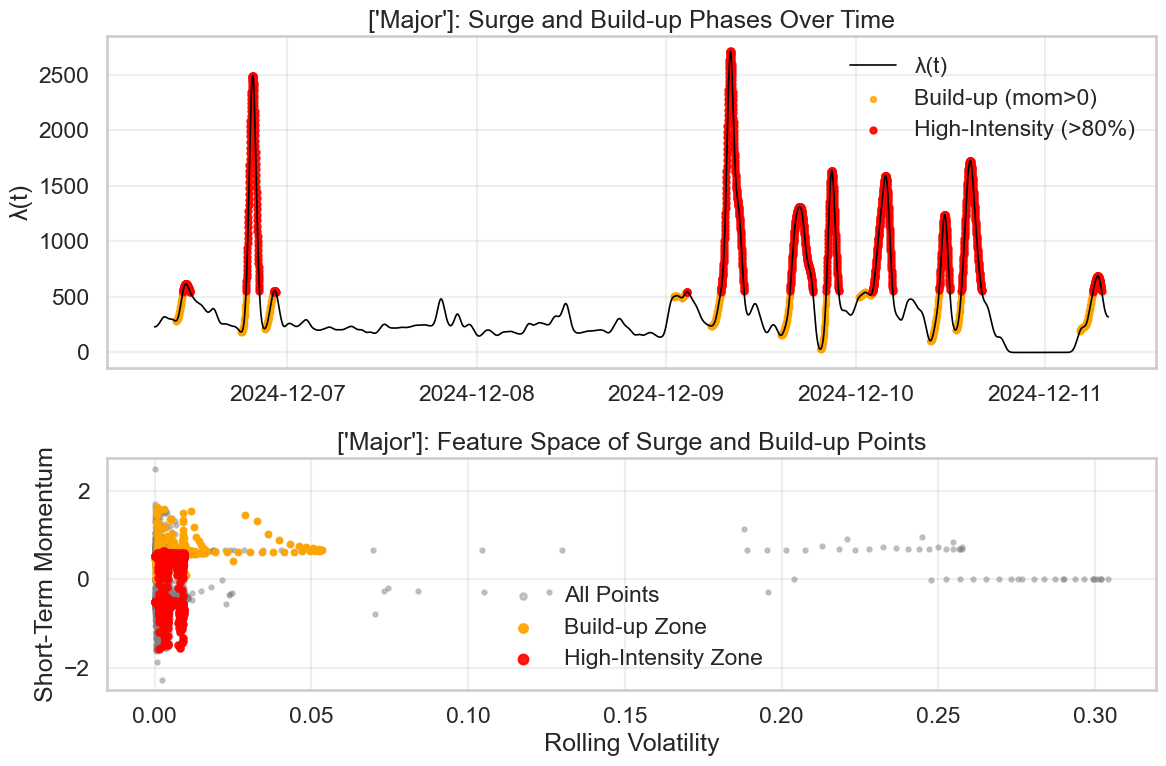}
    \caption{
    Regime overlays for the \textbf{Major} stratum. 
    The top panel shows the alert intensity $\lambda(t)$ colored by inferred regimes (baseline intensity, buildup, surge), 
    while the bottom scatter plot illustrates the corresponding regions in the volatility–momentum plane. 
    Together, these visualizations highlight transitions from stable to escalated attack behavior.
    }
    \label{fig:regime_overlay}
\end{figure*}

\subsection{Feature Engineering and Temporal Characterization}
To quantify local variations in attack arrival dynamics, we derive two interpretable temporal features from the per-minute alert intensity series $\lambda(t)$.
\paragraph{Short-Term Momentum.}
To quantify directional shifts in alert intensity, we define a standardized short-term momentum feature based on the first difference of the smoothed intensity series. Specifically,
\[m(t) = \frac{\lambda(t) - \lambda(t-1)}{\sigma_w(t) + \epsilon}\]
where $\sigma_w(t)$ is the rolling standard deviation of $\lambda(t)$ computed over a $w$-minute window (here, $w=5$), and $\epsilon$ is a small constant to prevent division by zero. This formulation standardizes the rate of change in alert intensity, making momentum dimensionless and comparable across strata. Positive values indicate short-term acceleration, while negative values reflect decline or stabilization.

\paragraph{Rolling Volatility.}
Volatility captures the dispersion of momentum over time, serving as a measure of local uncertainty or instability in attack arrivals. 
It is computed as:
\[
v(t) = \sqrt{\frac{1}{w}\sum_{i=t-w+1}^{t}\left(m(i) - \bar{m}_w\right)^2}
\]
where $w$ denotes the rolling window length (here, $w = 5$ bins) and $\bar{m}_w$ is the mean momentum over that window.

\subsection{Exploratory Analysis of the Volatility–Momentum feature-plane}

To understand the short-term dynamics of IDS alert streams and identify potential precursors to high-intensity intervals, we examine the joint distribution of volatility and momentum derived from the smoothed intensity process. Plotting each time point in the two-dimensional momentum-volailty plane reveals clear structural patterns that are not visible in the raw alert counts.

The majority of points occupy a stable low-variance region characterized by low volatility and weak or negative momentum. These periods correspond to routine background activity where alert arrivals fluctuate minimally around a local mean. We also identify a distinct “buildup” zone in which momentum is consistently positive while volatility remains moderate. Points in this region frequently precede transitions into higher-intensity states, suggesting that sustained upward pressure in the arrival rate acts as an early indicator of escalation. Finally, a sparse but well-separated cluster appears in the high-volatility, high-momentum region; these points align with empirically identified extreme-intensity events and correspond to pronounced surges in alert load. These regimes are shown in Figure \ref{fig:regime_overlay}.

These geometric patterns imply that volatility and momentum jointly encode meaningful temporal structure: volatility reflects instability and dispersion in the arrival process, while momentum captures directional acceleration. Their combination provides a compact and interpretable representation of short-term dynamics. Importantly, these regions emerge consistently across Suricata severity strata, although the density and boundaries differ across classes, reflecting heterogeneous operational behaviors. This analysis motivates the use of volatility–momentum features as predictive signals for forecasting tail-risk escalation in the IDS alert stream.

\section{Extreme-Regime Forecasting Framework}

\subsection{Problem Formulation: Tail-Risk Prediction}
\label{sec:problem}

Let $\{N(t): t \ge 0\}$ denote the IDS alert arrival process, modeled as a counting process adapted to a natural filtration $\{\mathcal{F}_t\}$ representing all information available up to time $t$. Let $\lambda(t)$ denote a smoothed estimate of the instantaneous alert intensity derived from $N(t)$, as defined in Section~3.

Our forecasting objective is to determine whether the next interval will enter an extreme-intensity regime defined by the empirical 95th percentile of $\lambda_t$. Formally, we define the binary prediction target
\[
y_t = \mathbf{1}\{\lambda_{t+1} \geq P_{95}(\lambda)\}.
\]

This formulation treats extreme alert surges as tail events and reframes escalation prediction as estimating the conditional probability
\[
\mathbb{P}(y_t = 1 \mid \mathcal{F}_t),
\]
that is, the probability of entering an extreme-intensity regime given all information available at time $t$.

The setup provides a direct operational analogue to value-at-risk (VaR) forecasting in finance and extreme-event prediction in nonlinear dynamical systems, bridging IDS telemetry with established models of tail-risk behavior. While the 95th percentile is used throughout this study as a representative high-risk operating point, the formulation naturally generalizes to alternative quantile thresholds reflecting different operational risk tolerances.

\subsection{Predictive Modeling and Evaluation}

To map temporal features to tail-risk probabilities, we adopt gradient-boosted decision trees (XGBoost). Gradient boosting is widely regarded as state-of-the-art for structured, tabular prediction problems due to its ability to capture nonlinear interactions, robustness to heterogeneous feature scales, and strong generalization performance \cite{chen2016xgboost}. XGBoost in particular has been successfully applied across domains such as anomaly detection, financial time-series forecasting, and high-dimensional risk modeling, making it well-suited for IDS telemetry with complex but low-dimensional feature interactions.

We train the model using the feature vector
\[
x_t = (\lambda_t, m_t, v_t),
\]
and learn the mapping
\[
f(x_t) \rightarrow \hat{y}_t,
\]
where $f(\cdot)$ outputs estimate that the event count at time t is in the 95th percentile $t$.

The binary target label is defined by the following indicator function:
\[
y_t =
\begin{cases}
1, & \text{if }\displaystyle \max_{\tau \in \{t+1, \dots, t+30\}} \lambda_{\tau} \;\ge\; q_{0.95}, \\[8pt]
0, & \text{otherwise}.
\end{cases}
\]

Thus, a positive label denotes that the system will enter an extreme-volume regime within the next 30 minutes. This formulation yields a realistic early-warning objective while ensuring strict temporal causality, since only features observable at time $t$ are used to predict future exceedances. To account for heterogeneity in alert behavior, we train separate models for each Suricata severity class.

We evaluate forecasting performance using a chronological split (70\% training, 30\% testing) to preserve temporal ordering and prevent future information leakage. Classification metrics include accuracy, precision, recall, F$_1$-score, ROC--AUC, and confusion matrices computed separately for each severity stratum.

The ablation study evaluates four feature configurations: intensity only $(\lambda_t)$, intensity with momentum $(\lambda_t, m_t)$, intensity with volatility $(\lambda_t, v_t)$, and the full feature set $(\lambda_t, m_t, v_t)$. Overall, the ablation results empirically validate our feature-engineering approach and reinforce the measurement-driven insights from our exploratory analysis of the arrival process.

\subsection{Experiments}

We conduct a series of experiments designed to evaluate the predictive accuracy, robustness, and interpretability of the proposed extreme-regime forecasting framework. All experiments are performed separately for each Suricata severity stratum to account for heterogeneity in alert arrival dynamics.

\paragraph{Extreme-regime forecasting.}
Our primary experiment evaluates the ability of the model to forecast entry into extreme-intensity regimes. For each severity stratum, we train a supervised classifier to predict whether the alert intensity will exceed the empirical 95th percentile within the next 30-minute horizon. Models are trained using a chronological split (70\% training, 30\% testing) to preserve temporal causality. Performance is assessed using standard classification metrics, including accuracy, precision, recall, F$_1$-score, ROC--AUC, and confusion matrices.

\paragraph{Feature ablation.}
To quantify the contribution of each temporal feature, we conduct an ablation study over four feature configurations: intensity only $(\lambda_t)$, intensity with momentum $(\lambda_t, m_t)$, intensity with volatility $(\lambda_t, v_t)$, and the full feature set $(\lambda_t, m_t, v_t)$. This experiment isolates the predictive power of volatility and momentum beyond the baseline intensity signal.

\paragraph{Interpretability and visualization.}
To examine how the model partitions temporal dynamics into risk regimes, we provide a proof-of-concept  visualization of the learned decision boundaries in the volatility–momentum feature plane. In addition, we generate operational risk snapshots showing predicted tail-risk scores across severity strata at representative time points, illustrating how the model can be used for early-warning and triage in practice.

\subsection{Temporal Data Leakage Prevention}
\label{sec:leakage}

To ensure a valid evaluation of temporal forecasting performance, we implement strict controls to prevent information leakage between training and test data. All design choices are explicitly structured to preserve temporal causality and to avoid contamination arising from temporal dependence.

\paragraph{Chronological split.}
We employ a forward-chaining split in which the first 70\% of the time series (December 6, 2024–January 15, 2025) is used for training, and the remaining 30\% (January 16–February 2, 2025) is reserved for testing. This chronological partition ensures that the model is trained exclusively on past data and never observes future information during training.

\paragraph{Feature engineering boundaries.}
All temporal features, including rolling volatility and momentum, are computed using strictly backward-looking windows. For predictions at time $t$, feature values are derived solely from observations in the interval $[t-w, t-1]$, where $w$ denotes the window size. We explicitly verify that no forward-looking information is incorporated into feature computation at any stage of training or evaluation.

\paragraph{Percentile threshold computation.}
The extreme-event threshold $P_{95}(\lambda)$ is computed exclusively from the training data and held fixed during evaluation on the test set. This prevents implicit leakage whereby the model could adapt to the distribution of future observations through dynamic thresholding.

\paragraph{Target label construction and data purging.}
For each training instance at time $t$, the binary target label is defined as
\[
y_t = \mathbf{1}\left\{\max_{\tau \in [t+1,\, t+30]} \lambda_\tau \ge P_{95}(\lambda)\right\},
\]
which depends only on future observations within the training period. To further mitigate temporal dependence across the train–test boundary, all observations within 30 minutes of the split point are removed from both the training and test sets. This purging step ensures that no partially observed future outcomes or correlated windows span the boundary, yielding a conservative and leakage-free evaluation protocol.

\section{Results and analysis}

\subsection{Predictive Performance across Strata}

The extreme-regime forecasting model achieves strong predictive performance across all severity strata, with accuracy exceeding 95\% when using the complete feature set (intensity, volatility, and momentum). Table~\ref{tab:strata-compact} presents the baseline performance metrics, while our ablation study reveals critical insights about feature importance.

The ablation analysis demonstrates that \textbf{intensity ($\lambda_t$) is the dominant predictive feature}. When intensity is removed (keeping only volatility and momentum), model performance degrades catastrophically across all strata---F1 scores plummet to near-zero for most categories. For instance, the Minor stratum's F1 score drops from 0.934 to 0.024 without intensity, while the Unknown stratum falls from 0.867 to 0.003.

In contrast, removing volatility or momentum individually causes moderate but recoverable performance degradation. The Minor stratum maintains an F1 of 0.162 without volatility, and the Critical stratum achieves 0.451 without volatility, suggesting these features provide complementary but non-essential information. This hierarchical importance (intensity $\gg$ volatility $>$ momentum) indicates that while temporal dynamics enhance prediction, the raw alert rate remains the primary signal for extreme-event forecasting.

Importantly, the resultant effect of combining all three features is most pronounced for high-severity strata. The Critical stratum shows a 97\% improvement in F1 score when adding temporal features to intensity alone (from 0.451 to 0.890), while lower-severity strata exhibit more modest gains. This suggests that sophisticated attacks exhibit richer temporal signatures that benefit from multi-feature characterization.

The Unknown stratum's unique behavior, maintaining relatively high AUC (>0.98) despite poor F1 scores in ablated models, indicates that while the model can distinguish extreme from normal regimes, precise threshold calibration becomes critical when temporal features are absent. This reinforces our design choice to engineer interpretable temporal features rather than relying solely on raw intensity.

\subsection{Operational Deployment Visualization
}
\begin{figure*}[t]
    \centering
    \includegraphics[width=0.8\linewidth]{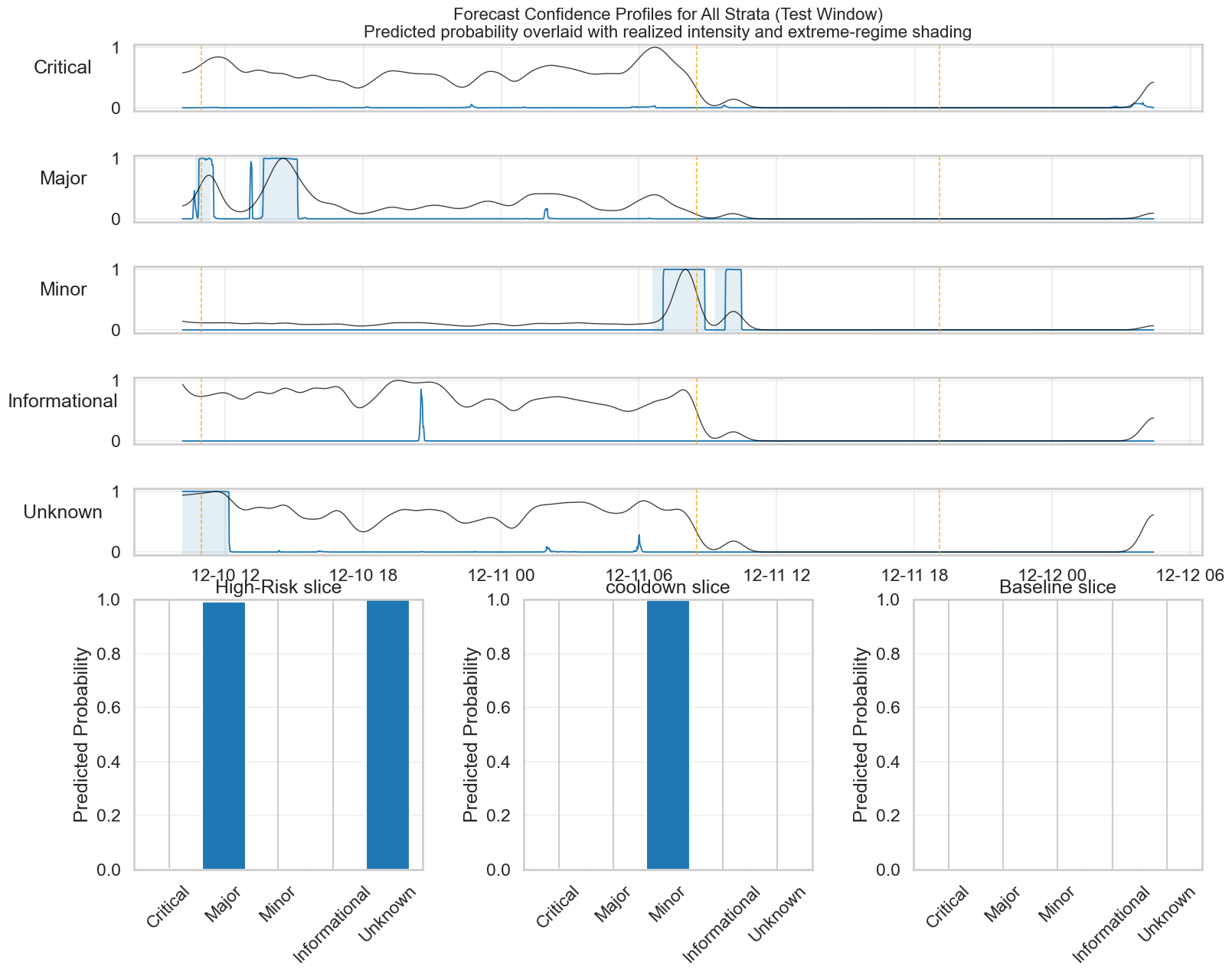}
    \caption{\textbf{Proof-of-concept visualization of forecasted tail risk and cross-stratum risk comparison.}
\textit{Top panels:} For each IDS alert stratum, the figure overlays the model-predicted probability of entering an extreme-intensity regime (blue) with the realized alert intensity (black, normalized for visualization). Shaded regions indicate time intervals labeled as extreme regimes under the 95th-percentile definition. Vertical dashed lines denote three selected time slices used for cross-sectional analysis.
\textit{Bottom panels:} Predicted tail-risk probabilities across all alert strata at the corresponding time slices, ordered from \textbf{high-intensity}, \textbf{cooldown}, to \textbf{baseline} operational states. Together, this proof-of-concept visualization demonstrates how temporal risk forecasts can simultaneously convey within-stratum escalation dynamics and relative risk distribution across strata at key operational moments.}
    \label{fig:confidence_viz}
\end{figure*}

Figure \ref{fig:confidence_viz} presents our proposed risk visualization showing confidence scores across all severity strata at three different time points. The top panel captures a baseline period where XGBoost confidence remains low across all strata (Unknown, Critical, Informational, Major, Minor), indicating normal operations. The middle panel shows a buildup phase where certain strata exhibit elevated confidence scores, signaling potential escalation in those specific alert categories. The bottom panel represents an active surge event where multiple strata show high confidence predictions simultaneously, indicating coordinated or widespread attack activity. This multi-strata view enables operators to quickly identify which attack types are escalating. For instance, high confidence in the Critical stratum but low confidence in Minor suggests targeted sophisticated attacks rather than broad scanning activity. The bar plot design provides an at-a-glance risk dashboard that complements traditional time-series views by explicitly showing per-stratum risk assessment rather than aggregated alert volumes.

\begin{table}[t]
\centering
\small
\begin{tabular}{lccccc}
\toprule
\textbf{Stratum} &
\textbf{Accuracy} &
\textbf{ROC AUC} &
\textbf{Precision} &
\textbf{Recall} &
\textbf{F1} \\
\midrule
Unknown & 0.9587 & 0.9968 & 0.8949 & 0.6926 & 0.7410 \\
Critical    & 0.9930 & 0.9987 & 0.9352 & 0.9272 & 0.9312 \\
Informational & 0.9897 & 0.9992 & 0.9329 & 0.9739 & 0.9530 \\
Major       & 0.9961 & 0.9997 & 0.9678 & 0.8864 & 0.9253 \\
Minor       & 0.9964 & 0.9995 & 0.9545 & 0.9850 & 0.9696 \\
\bottomrule
\end{tabular}
\caption{
Performance of the extreme-regime forecasting model across alert strata.
Metrics reflect positive-class (high-risk) precision, recall, and F1.
The ``Unknown'' category aggregates two smaller severity groups.
}
\label{tab:strata-compact}
\end{table}

\begin{table}[t]
\centering

\begin{tabular}{lccc}
\toprule
\textbf{Severity} &
\textbf{Full $(\lambda,m,v)$} &
\textbf{-- Volatility} &
\textbf{-- Momentum} \\
\midrule
Unknown (High $\lambda$) & 0.987 / 0.867 & 0.420 / 0.033 & 0.418 / 0.003 \\
Critical                 & 0.992 / 0.890 & 0.864 / 0.451 & 0.859 / 0.353 \\
Informational            & 0.997 / 0.935 & 0.946 / 0.518 & 0.925 / 0.577 \\
Major                    & 0.999 / 0.891 & 0.914 / 0.354 & 0.882 / 0.357 \\
Minor                    & 0.994 / 0.934 & 0.637 / 0.162 & 0.609 / 0.025 \\
Unknown (Low $\lambda$)  & 0.996 / 0.467 & 0.959 / 0.000 & 0.970 / 0.165 \\
\bottomrule
\end{tabular}
\caption{Ablation study across IDS severity strata. We report ROC--AUC and F$_1$ score for the full feature set $(\lambda, m, v)$ and for models with volatility or momentum removed. Dropping volatility or momentum results in substantial degradation, indicating that predictive performance is driven by higher-order temporal dynamics rather than static intensity alone.}
\label{tab:ablation}
\end{table}

\begin{figure}[t]
    \centering
    \includegraphics[width=0.999\linewidth]{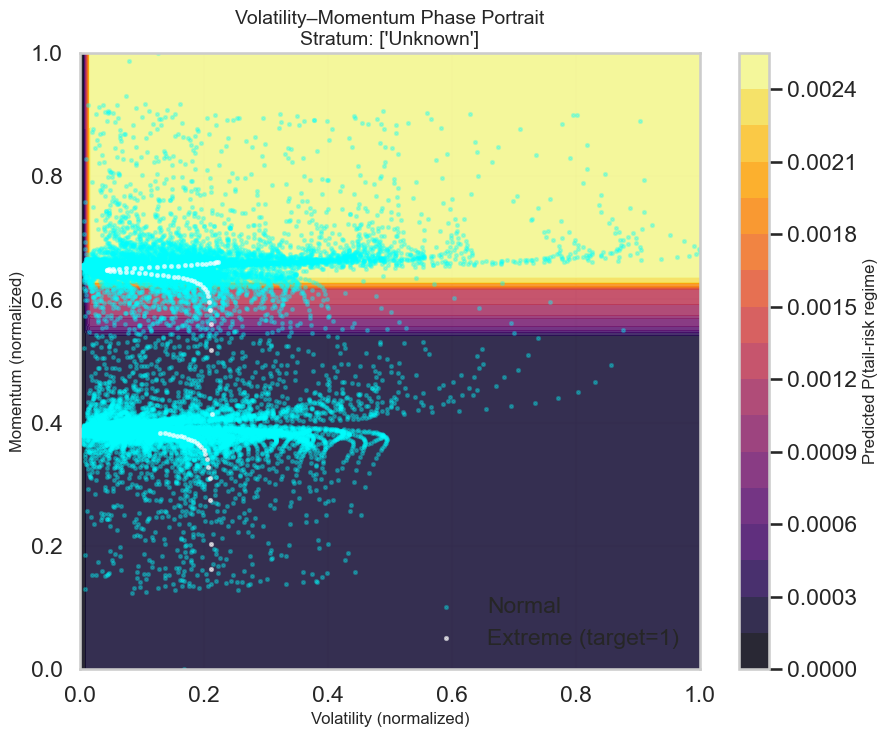}
    \caption{\textbf{Volatility–momentum phase portrait of predicted tail risk.}
The figure visualizes the decision surface learned by the XGBoost model in the volatility–momentum feature space. Each point represents a one-minute observation, while the background shading indicates the predicted probability of transitioning into an extreme-intensity regime. The color bar reports the uncalibrated probability values produced by the classifier.}
    \label{fig:xgboost_learns}

\end{figure}
\subsection{Interpreting Learned Risk Surfaces}

Figure~\ref{fig:xgboost_learns} presents the learned decision boundaries in the volatility-momentum feature space, revealing how XGBoost partitions temporal dynamics into risk regimes. The visualization demonstrates three distinct risk zones: low risk, transitional, and surge zone.

%The time required to identify the zones is ADDEDD 30 minutes in 4.1 

\textbf{Low-risk baseline} (blue or darkest grey lowest region): Characterized by low volatility (<0.8) and negative to neutral momentum, representing stable background activity. 
\textbf{Transitional buildup} (purple-orange gradient or four gray stripes second darkest): A region with positive momentum (0.2-0.6) and moderate volatility (0.2-0.4), serving as a gateway to high-intensity states.
\textbf{High intensity zone} (yellow-orange gradient): A wedge-shaped region with positive momentum (0.2-0.6) and moderate volatility (0.2-0.4), serving as a gateway to high-risk states.  The upper-right quadrant has both high volatility (>0.4) and strong positive momentum (>0.6), where extreme-event probability exceeds 80\%. This region is sparsely populated (<2\% of observations) but captures the vast majority of true extreme events.

Superimposed are data points, each representing one minute of data with the corresponding features. These appear light blue (or the second lightest gray in grayscale). 

The learned boundaries exhibit notable non-linearity which indicates that the model requires both elevated volatility and positive momentum to predict imminent escalation, rather than treating these features independently. The inclusion of both temporal features provides a stronger indicator than either feature alone.

\section{discussion}
\label{disc}

\subsection{Methodological Implications}
Methodologically, the results demonstrate that volatility and momentum features—often used to detect short-horizon regime shifts in other domains are informative in IDS telemetry. IDS alert streams reflect a mixture of activity types, including automated scanning, targeted intrusions, and benign background traffic, which favors flexible, nonparametric models. In line with the strong empirical performance of gradient-boosted trees in other structured prediction settings~\cite{chen2016xgboost}, our evaluation shows that gradient-boosted decision trees perform robustly across severity strata without requiring explicit distributional assumptions or regime identification.

\subsection{Operational Significance}
From a systems perspective, the proposed framework addresses a mismatch between IDS design objectives and SOC operational failure modes. IDS systems optimize per-alert detection, yet operational breakdowns are often driven by temporal overload effects, queue buildup, context switching, and degraded prioritization under sustained alert pressure. By forecasting entry into extreme-intensity regimes within a short horizon (30 minutes) with high discriminative accuracy across strata (Table~\ref{tab:strata-compact}), our approach enables proactive resource planning, such as pre-allocation of analyst attention, temporary adjustment of filtering thresholds, or short-term scaling of supporting infrastructure.

The stratified analysis further suggests that not all escalations carry equal operational meaning. High-severity strata show larger marginal gains from temporal features, indicating that these streams may contain richer short-horizon precursors than purely volumetric alert floods (Table \ref{tab:ablation}). This supports adaptive monitoring strategies where temporal forecasting is prioritized for high-value alert streams rather than uniformly applied across all traffic.

\subsection{Positioning Among Existing Approaches}
The proposed extreme-regime forecasting formulation occupies a distinct and complementary position within the security analytics landscape. Unlike anomaly detection, which focuses on deviations from historical baselines, our approach predicts entry into predefined high-risk operational states. Unlike alert correlation, which reduces redundancy after alerts are generated, we provide anticipatory signals that can support operational planning.

\subsection{Temporal Structure and Extreme-Regime Dynamics}

By framing the target as whether the alert-intensity process will cross a high quantile within a short prediction horizon, we obtain an operationally meaningful signal for planning and managing human resource pipelines in a security operations center (SOC). The model does not simply learn static thresholds, a common limitation of threshold-based approaches, but instead captures the underlying dynamics of alert arrivals. By leveraging higher-order temporal information, such as momentum and volatility, the model learns patterns that generalize across operating conditions and evolving threat environments.

\section{Limitations and Future Work}
This section focuses exclusively on methodological limitations and concrete avenues for extending the present work, complementing the interpretive discussion in Section~\ref{disc}.

While the empirical findings demonstrate that IDS telemetry contains measurable precursors to extreme-intensity events, the present study remains bounded by several methodological limitations that open important avenues for future work. First, our modeling scope is intentionally narrow: we rely on a single, widely adopted supervised learning method (XGBoost) to ensure interpretability and deployment realism. This choice allows us to characterize how volatility, momentum, and intensity combine to form predictive signatures, but it also restricts the breadth of representational capacity explored. More expressive architectures, such as deep temporal models, state-space formulations, quantile transformer networks, and recurrent Hawkes processes, may uncover structures that tree ensembles cannot. A systematic comparison across model families is therefore an important direction for extension.

Second, the study evaluates prediction performance only for a single extreme-event definition. The model as currently tuned targets exceedance of the 95th percentile of the intensity distribution. This threshold was selected as a representative high-volume operating point, corresponding to alert-flood conditions that are known to induce analyst overload and alert fatigue in SOC workflows \cite{SANSSOC2019}. Future work should broaden the analysis to encompass multiple quantile thresholds (e.g., 90\%, 97.5\%, 99\%), continuous quantile-regression approaches, or full distributional forecasting. In practice, an organization should be able to tune the model to align with its internal risk tolerance. Such extensions would allow SOC operators to modulate alerting behavior based on varying risk appetites rather than relying on a single hard cutoff.

Third, while the volatility and momentum features are grounded in practices widely used in econometrics and market-microstructure research, we do not perform a full parameter-sensitivity study. A more exhaustive sensitivity analysis could reveal feature scales that are more appropriate for cybersecurity telemetry, which may exhibit temporal structure distinct from financial time series. %The smoothing bandwidths, volatility windows, and momentum horizons were selected for conceptual clarity rather than as the result of formal optimization. 

\begin{figure}[h]
    \centering
    \includegraphics[width=0.7\linewidth]{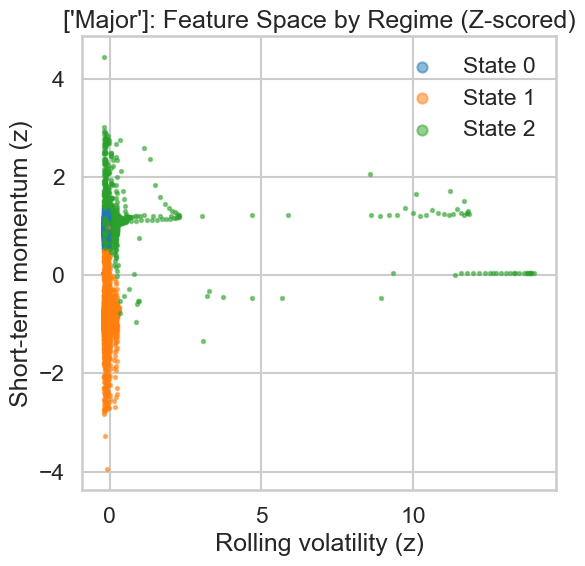}
    \caption{Failure of HMMs in generalizing  to clusters}
    \label{fig:hmmfail}

\end{figure}

Fourth, although escalation dynamics are inherently sequential, our attempts to model regime transitions using Gaussian Mixture Models and Hidden Markov Models were unsuccessful. These models were unable to learn stable latent states or generalize across strata. This suggests that IDS alert streams may require more flexible temporal clustering approaches. See Figure \ref{fig:hmmfail} to see how the fail to cluster in some of our initial experimentation. %, such as semi-Markov models\cite{yu2010hidden}, switching-Hawkes processes\cite{lewis2011nonparametric}, or neural state-space models \cite{kidger2020neural, rubanova2019latent} capable of capturing both excitation and decay patterns.

The present analysis captures only volumetric risk. Extreme-intensity forecasting focuses solely on fluctuations in alert counts, yet operational risk is multidimensional, encompassing adversary behavior, technique transitions, protocol characteristics, and the interaction of multiple strata. Integrating MITRE ATT\&CK trajectories, protocol-level distributions, or cross-stratum dependencies may yield a more holistic and operationally actionable picture of imminent escalation.

Finally, in terms of the presentation of the results, there is no question that the threshold method will scale. However, the method of presenting this may be ineffective.  We face the common challenge of other usable security researchers in that system analyst time is highly valuable. Thus, recruiting system analysts for a theoretical tool has proven challenging and is a future goal.

Together, these limitations reflect deliberate choices to constrain the scope of the work. The objective of the present study is not to optimize predictive performance, but rather to establish that IDS telemetry exhibits statistically measurable precursors to future high-intensity periods. 

In future work, we will intend to both deepen the analysis of IDS and examine if other patterns of attacks can be identified using extreme-regime forecasting methods. Additional IDS analysis from other organizations can provide depth to the analysis. For that reason, our model is available at \textit{anonymized until publication}. In terms of breath, other types of attacks can conceptually be modeled as extreme-regime forecasting challenges. Previous work has examined these in time series terms using temporal graph theory. It  been applied to identify, for example, BGP hijacks~\cite{ moriano2021using}, malware~\cite{ sleeman2020temporal}, and security policies~\cite{ bag2025enhancing}. We would welcome collaborators and data contributions to enable a broader comparison of the efficacy of extreme-regime forecasting with temporal graph methods.

\section{Conclusion}
This study presents initial evidence that volumetric patterns within IDS alerts exhibit a short-horizon structure that can be exploited for forecasting elevated-intensity periods. Further, the predictions are effective for each stratum o attacks. Rather than claiming the presence of universal “precursors,” our results show that a simple, interpretable feature set has the potential for early detection of increasing future attacks. Our core insight is that framing IDS alerts as a temporal-risk forecasting problem and constructing interpretable volatility–momentum–intensity features enables distinguishing those that will accelerate at a high level of accuracy. The strongest predictive signal originates from the local intensity trajectory itself, while volatility and momentum provide smaller but complementary contributions, particularly in strata with lower baseline alert rates.

These findings do not establish a general early-warning system, but they demonstrate that IDS streams contain useful short-range temporal information, enough to produce non-trivial predictive probabilities of entering extreme high-volume regimes. The cross-sectional confidence visualizations further illustrate how different strata respond differently at matched temporal snapshots, suggesting that SOC operators may benefit from stratified rather than aggregate forecasting.

Our study is intentionally scoped: we examine a single model family, a single extreme-event threshold, and only volumetric risk. As such, the contributions should be interpreted as proof-of-concept rather than definitive claims about predictive precursors. Nonetheless, the results open a promising path toward richer, interpretable IDS-driven risk forecasting frameworks.

\section{acknowledgements}

We thank Anonymous for their help; and thanks to the reviewers in advance.

\bibliographystyle{ACM-Reference-Format}
\bibliography{ref}
\end{document}